\documentclass[pdflatex,sn-basic,Numbered]{sn-jnl}
\usepackage{manyfoot}
\usepackage{xcolor}
\usepackage{amsmath,amssymb}

\title[All-dielectric vapor cells]{Batch-fabrication of all-dielectric vapor cells enabling optically addressed Rydberg atom electrometry}

\author*[1]{\fnm{Alexandra B.} \sur{Artusio-Glimpse}}\email{alexandra.artusio-glimpse@nist.gov}
\author[1,2]{\fnm{Adil} \sur{Meraki}}\email{adil.meraki@nist.gov}
\author[3]{\fnm{Hunter} \sur{Shillingburg}}\email{hfs5158@psu.edu}
\author[5]{\fnm{Guy} \sur{Lavallee}}\email{gpl107@psu.edu}
\author[5]{\fnm{Miao} \sur{Liu}}\email{Liumiao122@gmail.com}
\author[5]{\fnm{Chad} \sur{Eichfeld}}\email{cme133@psu.edu}
\author[1]{\fnm{Nikunjkumar} \sur{Prajapat}}\email{nikunjkumar.prajapati@nist.gov}
\author[1]{\fnm{Matthew T.} \sur{Simons}}\email{matthew.simons@nist.gov}
\author[4]{\fnm{Glenn} \sur{Holland}}\email{glenn.holland@nist.gov}
\author[1]{\fnm{Christopher L.} \sur{Holloway}}\email{christopher.holloway@nist.gov}
\author[4]{\fnm{Vladimir A.} \sur{Aksyuk}}\email{vladimir.aksyuk@nist.gov}
\author*[3,4,5]{\fnm{Daniel} \sur{Lopez}}\email{dlopez@psu.edu}

\affil[1]{\orgdiv{Electromagnetic Fields Group}, \orgname{National Institute of Standards and Technology}, \orgaddress{\street{325 Broadway}, \city{Boulder}, \postcode{80305}, \state{CO}, \country{US}}}
\affil[2]{\orgdiv{Physics Department}, \orgname{University of Colorado Boulder}, \orgaddress{\street{1234 Main St.}, \city{Boulder}, \postcode{80303}, \state{CO}, \country{US}}}
\affil[3]{\orgdiv{Electrical Engineering and Computer Science}, \orgname{Penn State University}, \orgaddress{\street{207 Electrical Engineering West}, \city{University Park}, \postcode{16802}, \state{PA}, \country{US}}}
\affil[4]{\orgdiv{Microsystems and Nanotechnology Division}, \orgname{National Institute of Standards and Technology}, \orgaddress{\street{100 Bureau Drive}, \city{Gaithersburg}, \postcode{20899}, \state{MD}, \country{US}}}
\affil[5]{\orgdiv{Materials Research Institute}, \orgname{Penn State University}, \orgaddress{\city{University Park}, \postcode{16802}, \state{PA}, \country{US}}}

\abstract{Millimeter-scale atomic vapor cells can be accurately and economically batch-fabricated by anodically bonding silicon and glass wafers, enabling the manufacturing of miniature atomic clocks and quantum sensors. However, silicon's high dielectric constant and conductive losses at millimeter wave frequencies limit its suitability for Rydberg-atom electrometry which enables highly sensitive electric-field measurements by exploiting the extreme polarizability of Rydberg states in alkali atoms. To address this, we present an all-glass wafer-level microfabrication process that eliminates silicon, creating hermetically sealed vapor cells that are stable over long timelines with embedded cesium dispensers. Femtosecond laser machining precisely defines the cell geometry, and laser-activated alkali loading ensures reliable filling. We demonstrate long-term vacuum stability and robust Rydberg excitation through electromagnetically induced transparency measurements. We then use these cells to measure a 34~GHz millimeter wave field resonant with the $58D_{5/2} \rightarrow 60P_{3/2}$ transition using Autler-Townes splitting and observe the expected linear dependence with field strength. This work demonstrates that the all-glass bonding approach offers a highly durable low-loss cell alternative for miniaturized millimeter wave and microwave quantum sensing, with the potential to flexibly incorporate a range of other dielectric and semiconductor materials and integrate with photonic and electronic technologies.}

\begin{document}
\maketitle
\section{Introduction}\label{introduction}

Miniaturized alkali-vapor cells have enabled a range of precision
measurement technologies, including global positioning systems,
telecommunications, chip-scale magnetometers, and more recently, broadly
tunable Rydberg-atom electric field sensors. Over the last two decades,
as these applications have diversified, a variety of fabrication
techniques have emerged, each optimized for a different operating
regime. The most prominent methods include glass-blown vapor cells at
the millimeter scale and anodically bonded glass-and-Si chip-scale vapor
cells, both of which allowing batch fabrication. Elsewhere,
heterogeneous integration of diverse material platforms, such as III-V
semiconductors and glass substrates, has shown great promise for
high-speed electronics and advanced integrated photonics. This
integration is often achieved through direct bonding, where cleaned,
polished, and activated surfaces are brought into contact and annealed
at modest temperatures to form strong covalent
bonds~\cite{Plol1999,Masteika2014}. As chip-scale vapor cells are integrated
with photonics systems~\cite{Hummon2018,Newman2019,Moody2022,Yulaev2024,Zektzer2024,Long2024}, new compatibility
challenges arise, driving the need for new cell materials, structure
geometries, and innovative fabrication processes.

Rydberg atom electrometry is a technique that uses the extreme
sensitivity of Rydberg atoms to electric fields for precise
measurements~\cite{Schlossberger2024}. Alkali atoms, excited to high
principal quantum numbers, exhibit exaggerated electromagnetic coupling,
making them ideal for optically detecting and measuring electric fields
with high accuracy across an extended and tunable range of frequencies.
Detection of over-the-air millimeter wave (30~GHz to 300~GHz) signals
using Rydberg atoms presents new challenges and opportunities for
optimizing the incident field interactions with the vapor cell
materials, structured at the millimeter scale comparable to the RF
wavelength. Combined with the need for photonic integration of the
optical fields for excitation and read-out of Rydberg atoms, we were
motivated to develop a direct bonding wafer-level fabrication process
for small vapor cells that offers broader material options. The stable
high vacuum hermetic seal critical to a wide range of cell applications
is stringently tested by the ground state and Rydberg state spectroscopy
measurements of Rydberg electrometry.

Glass-blown alkali cells are produced in sizes ranging from several
centimeter cylinder and spherical cells for laboratory spectroscopy down
to 3~mm to 5~mm spheres for satellite clocks and compact magnetometers
and gyroscopes. Below roughly 5~mm, the process leads to measurable
birefringence and surface curvature, imperfections that must be
mitigated by post-blow annealing or polarimetric
compensation~\cite{Balabas2006}. Hollow-core photonic-crystal fibers
and related silica waveguides offer a complementary route to
miniaturization. Alkali vapor can be loaded into 5~\(\mu m\) to
150~\(\mu m\) cores, giving centimeter-long interaction lengths and
large optical depths suitable for a range of sensing applications.
However, a practical consideration is that vapor fills the fiber
axially, so local density variations can introduce static field
gradients along the fiber length, an effect that must be managed in
high-precision Rydberg measurements~\cite{Epple2017,Haupl2022}.

Millimeter-scale vapor cells used in atomic
clocks~\cite{Knappe2004,Martinez2023}, optically pumped
magnetometers~\cite{Ma2024}, and absolute gravity
sensors~\cite{Li2023}, on the other hand, are accurately
batch-fabricated from silicon and borosilicate glass wafers using
optical lithography and anodic bonding~\cite{Wallis1969,Kitching2018}. However,
anodic bonding typically requires doped silicon, typical resistivity in
the 1~\(\Omega \cdot cm\) to 30~\(\Omega \cdot cm\)
range~\cite{Knowles2006}, a material with high dielectric permittivity
and conductive loss for radio frequency (RF)
fields~\cite{Hasegawa1971}. The resulting strong absorption,
scattering, and local perturbation of the incident RF fields is
undesirable for Rydberg atom-based RF measurements and may be
detrimental in other atomic quantum sensing scenarios.

Rydberg-atom-based measurements of microwave fields have been shown in
silicon-based anodically bonded vapor cells~\cite{Zhao2023,Giat2025_arxiv};
however, improved field sensitivity is expected with more RF-friendly
cell materials. Thinning the silicon walls is one option to minimize the
undesired field-structure interaction~\cite{Shi2024}.
Alternatively, sputtered thin-film silicon~\cite{Berthold2000} can be
used in anodic bonding with borosilicate or aluminosilicate glasses to
minimize the volume of silicon in the vapor cell. Subwavelength silicon
patterning has been proposed for engineering the dielectric constant to
minimize scattering of the RF field~\cite{Noaman2023,Pandiyan2024}. Such
dielectric structure engineering can be further optimized if a broader
choice of cell materials became available. High resistivity intrinsic
silicon can be bonded to glass using plasma surface
treatment~\cite{Howlader2010}. The most promising technique for
batch-manufacturing anodically bound vapor cells from high-resistivity
silicon involves the deposition of
Al\(_{2}\)O\(_{3}\)~\cite{Raghavan2024}, which acts as an adhesion
layer~\cite{Sahoo2017} and reduces the consumption of
alkali~\cite{Karlen2017}. But, regardless of the bonding technique
used, the high dielectric constant of intrinsic silicon at
\(\approx 11.7\)~\cite{Yang2019} compared to that of fused silica
at \(\approx 4\) and Borofloat 33 \cite{SchottBorofloat33}~at
\(\approx 4.5\)~\cite{Cano2024}, presents a much stronger spatially
dependent perturbation of the electric field measured by the atoms,
which is undesirable for many millimeter wave applications. Therefore,
all-glass vapor cells for these applications are preferred.

A small number of miniature glass vapor cells have been reported using
bonding techniques such as glass fusion~\cite{Kubler2010}, optical
contact bonding~\cite{Peyrot2019,Cutler2020}, edge
welding~\cite{Baluktsian2010}, epoxy glue~\cite{Daschner2012,Lucivero2022}, hot
wire cutting~\cite{Laliotis2022}, and even anodic bonding via a 200~nm
thick sputtered layer of silicon nitride~\cite{Daschner2014}. Although
these demonstrations of small, even micrometer-sized vapor cells prove
that such miniature vapor cells can host Rydberg atoms for a range of
applications, the fabrication schemes are not easily scalable to
batch-fabrication with the exception of the silicon nitride anodic
bonding method.

Here, we demonstrate the feasibility of using direct bonding for
wafer-scale manufacturing of high performance alkali-vapor cells,
expanding the toolkit for integrating atomic vapor with photonics,
lasers, electronics, and mechanical devices. By successfully fabricating
and filling glass vapor cells with long observed lifetimes, our work
broadens the range of materials available for mass manufacturing and
wafer-level integration of atomic vapor, paving the way for a new
generation of heterogeneously integrated systems. Borofloat 33
substrates are used because of the familiarity of this material in many
vapor cell systems and because of their low dielectric constant and low
loss in the RF domain. With this demonstrated microfabrication of
high-quality direct bonded cells using a commercial 150~mm wafer bonding
tool, we overcome current fabrication limitations and pave the way for
many more materials to be used in miniature atomic vapor systems.

We detail electromagnetically induced transparency (EIT) measurements
using cesium (Cs) Rydberg atoms, carried out repeatedly over the period
of 23 months, indicating good performance over a long period.
Comparisons of cells from different wafer batches suggests reliability
of this manufacturing process. We carry out Rydberg atom electrometry of
a 34~GHz electric field using the 58D\(_{5/2} \rightarrow\)60P\(_{3/2}\)
resonant transition as a test case to demonstrate the applicability of
this cell manufacturing technique for the detection of millimeter wave
fields. Using wafer-level femtosecond laser machining, we accurately
create intricate cavity geometries in all-dielectric cells that can be
tailored to minimize or custom-engineer near- and far-field scattering
of the RF field, while avoiding deleterious absorption, resulting in a
tailorable and more robustly predictable relationship between the RF
fields outside and inside the vapor cell over a wide range of
frequencies. The presented cells with approximately 1.9~cm x 1.3~cm
cavities have been engineered specifically for the future goal of
performing sub-wavelength measurements and mapping of a millimeter wave
field using the Rydberg atoms at multiple spatial locations
simultaneously.

\begin{figure}[t]
  \centering
  \includegraphics[width=\textwidth]{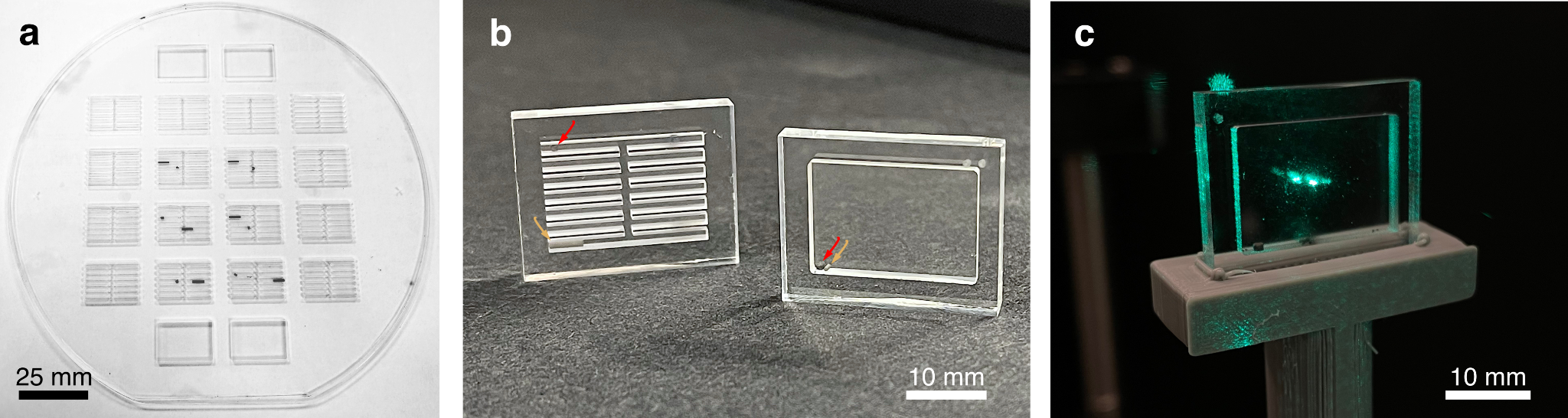}
  \caption{Photographs of (a) a bonded 150~mm diameter wafer having 20
vapor cells with different geometries, six of which in this case contain
a cesium dispenser and a getter, (b) two filled vapor cells - a
supported version and an open version (red arrows point to Cs dispensers
and orange arrows point to getters), and (c) an open cell shown with the
Rydberg lasers propagating through it.}
  \label{fig:vaporCellPictures}
\end{figure}

\section{Results}\label{results}

\subsection{Fabrication and Filling
Process}\label{fabrication-and-filling-process}

Our wafer-level vapor cells are fabricated by bonding a triple-stack of
borosilicate wafers to form a full 150~mm wafer of twenty vapor cells,
as shown in Figure~\ref{fig:vaporCellPictures}(a-c). Bonded
stacks are diced into individual vapor cell die with nominal outer
dimensions of \(25\ mm\  \times \ 20\ mm\). The thickness of the middle
wafer, which defines the length of the optical path within the vapor
cell, is 2~mm and the thickness of each of the window wafers is 1~mm.
The internal cavity volume of each vapor cell is machined through the
middle wafer prior to bonding using femtosecond laser pulses and
potassium hydroxide (KOH) etching, resulting in a variety of internal
cell geometries tailored to the application needs. The KOH etch is
highly selective to the denatured glass areas defined by femtosecond
laser exposure, preserving the surface quality of the bond interface
without need for additional polishing. Throughout this paper, we discuss
measurements with two internal cell geometries, supported and open, as
pictured in Figure~\ref{fig:vaporCellPictures}(b) and discussed
in more detail in Section~\ref{sec:vapor-cell-geometry-details}.
Notably, these cells come from different wafer batches, the supported
cell bonded in January 2023 and the open cell bonded in September 2024,
which highlights the reliability and repeatability of this manufacturing
process.

Further miniaturization is feasible with this wafer-level process; in
practice the minimum lateral footprint can be pushed down toward the
alkali source pill size. The main limiting factor is therefore not the
bonding/etching, but reliable handling, placement, and sealing of very
small source pieces under vacuum while maintaining clean, hermetic
interfaces.

\begin{figure}[t]
  \centering
  \includegraphics[width=0.55\textwidth]{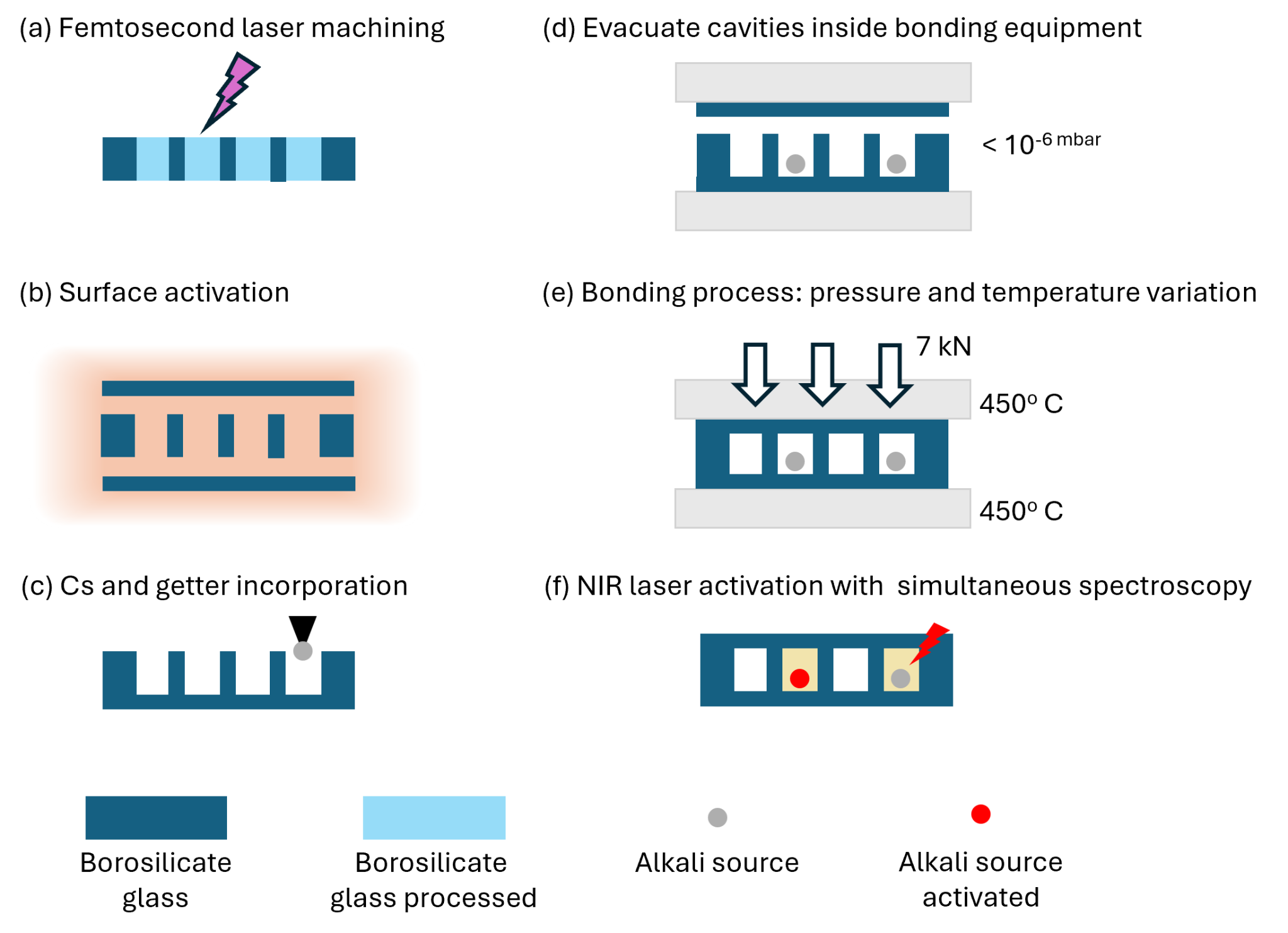}
  \caption{Vapor cell fabrication process steps: (a) femtosecond (fs)
laser machining of the 2~mm thick glass wafer, (b) cleaning and surface
activation of glass wafers before bonding of the preform as depicted in
(e), (c) placement of the alkali dispensers and non-evaporable getters,
(d) introduction into bonding equipment and evacuation of the bonding
chamber down to \(10^{- 6}\)~mbar, (e) application of pressure at
elevated temperature for 20 hours, and (f) activation of the alkali
source pills by near-infrared (NIR) laser heating. Wafer dicing to
separate each vapor cell is performed either prior to or just after step
(f).}
  \label{fig:fabSteps}
\end{figure}

The wafer fabrication sequence is depicted in
Figure~\ref{fig:fabSteps}. Following the machining of the glass
frame (middle wafer), the machined wafer and both the top and bottom
window wafers are placed in a 75\(\ ^{\circ}\) sulfuric acid/hydrogen
peroxide Nanostrip~(n.d.) bath for one hour, which cleans and activates
the surfaces of the wafers. After removal from the chemical bath, the
wafers are rinsed with deionized water and dried using N\(_{2}\) in a
spin-rinse dryer. After drying, the bottom window wafer and the machined
middle wafer are brought into contact to form a partially sealed cell (a
preform). Next, placement of a Cs dispenser (Cs-56Zr-11Al) and a
non-evaporable getter (NEG) within each cell cavity is performed using a
vacuum pen to minimize particle deposition on the bond surface. The NEG
is added for additional background gas pumping, if needed. Immediately
after placing the Cs dispensers and getters in the cells, the partially
sealed cells and the top window wafer are loaded into the bonding
equipment. The time between the clean step and wafer contact for bonding
must be 30 minutes or less for a successful bond to occur.

The bonding process starts by evacuating the system to a base pressure
of \(< \ 10^{- 6}\ mbar\) while at the same time keeping the bonding
substrates separated to ensure that the completed vapor cells will be
sufficiently evacuated after bond. Once the appropriate vacuum level is
achieved, the wafers are then brought together and heated to a
temperature between 450~\(^{\circ}\)C to 500~\(^{\circ}\)C with a force
of \(\approx 7\ kN\) applied to the wafers. After holding these
conditions for 20 hours, the temperature is then set to
20~\(^{\circ}\)C, and the applied pressure is turned off so that the
wafers are allowed to cool and relax naturally.

\begin{figure}[t]
  \centering
  \includegraphics[width=\textwidth]{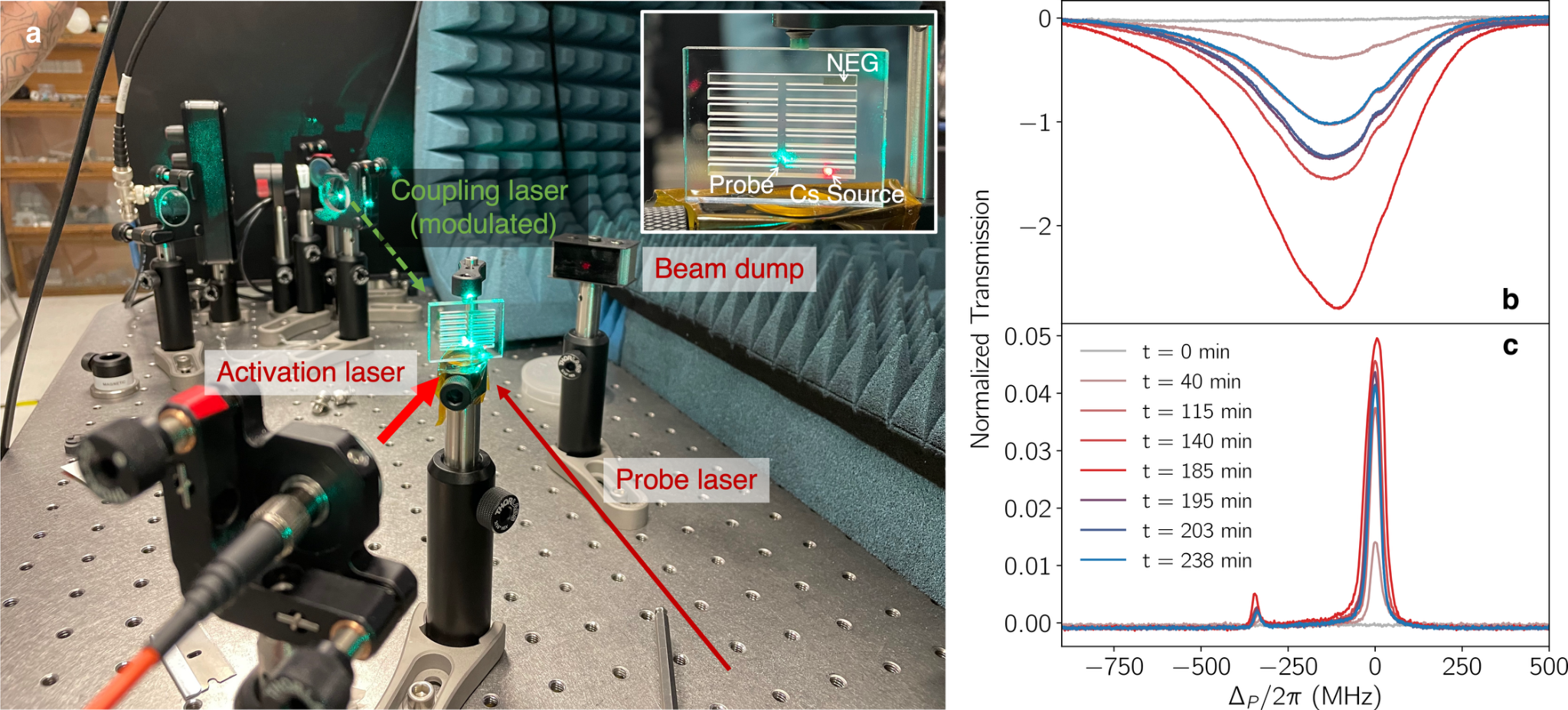}
  \caption{(a) Dispenser activation setup including inset picture of a
wafer-processed glass vapor cell (red light on Cs dispenser is guide
beam coaligned with the \(\approx 1\)~W, 980~nm activation laser). Plots
on right show simultaneous (b) absorption and (c) 50D\(_{5/2}\) EIT
measurements over time as the Cs dispenser was activated. Activation
laser turned off at \(t = 201\ min\) after which the vapor cooled and
stabilized. Data has been smoothed for improved visibility.}
  \label{fig:activation}
\end{figure}

Upon completion of the bonding, each vapor cell is diced out of the
wafer stack and filled by laser-heated activation of the Cs dispensers;
the order between filling and dicing is interchangeable. Laser-heated
activation of the getters was completed in some of the vapor cells, but
it was not found to have a measurable impact on the measured spectra,
suggesting evacuation in the bonding tool was sufficient (see
Section~\ref{sec:eit-spectra-and-rf-electrometry} for further
analysis of the background pressure in one of these vapor cells). During
laser activation, a two-photon Rydberg EIT signal is monitored while
scanning the probe laser over the D2 Doppler absorption profile. See
Figure~\ref{fig:activation} for example data taken from one of
these test cells monitoring the 50D\(_{3/2}\) and 50D\(_{5/2}\) Rydberg
states. We used Cs dispensers in this proof-of-concept study because
they are common alkali sources in microfabricated vapor cells. However,
the Zr-Al scaffolding left behind as a pill (also a getter) after Cs
release may perturb incident RF fields. Future iterations will explore
alternative filling processes, such as using pure alkali metals or
methods that separate the alkali source from the final vapor cell.

\begin{figure}[t]
  \centering
  \includegraphics[width=\textwidth]{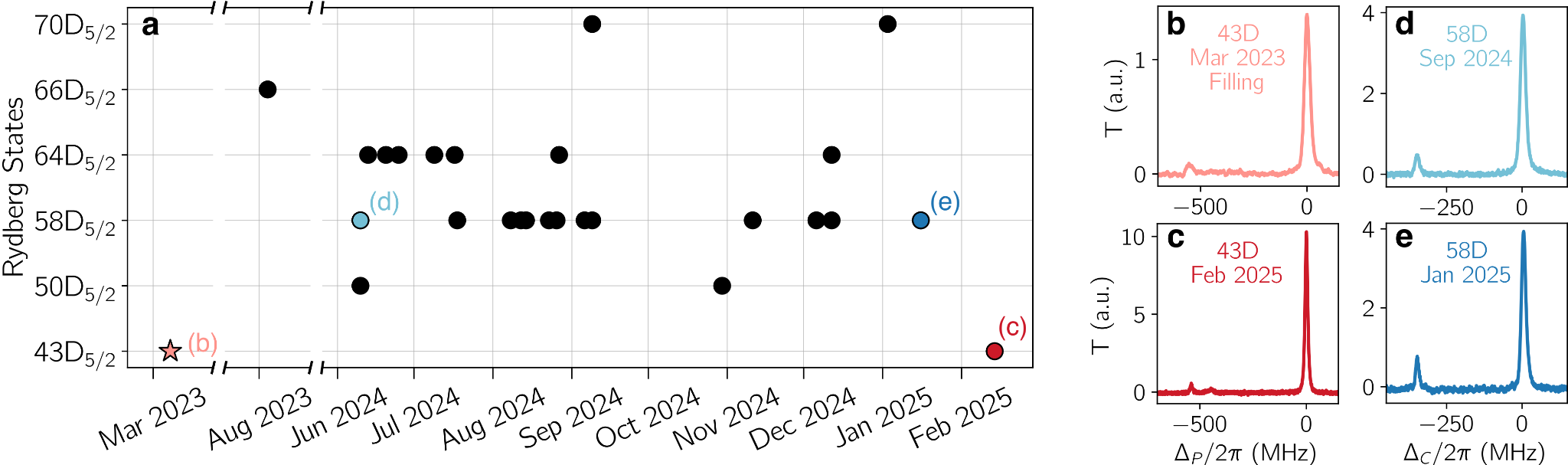}
  \caption{(a) Dates of Cs filling by laser heated activation (star) and
all Rydberg EIT spectroscopy measurements collected with one supported
vapor cell (circles) from a wafer lot bonded in Jan 2023. (b-e)
Selection of recorded EIT spectra corresponding to labeled markers in
(a).}
  \label{fig:timeline}
\end{figure}

The first batch of vapor cells using this direct bonding process and
laser-heated dispenser activation were bonded on 25 January 2023, the
activation study was completed on 6 March 2023, and regular measurements
of Rydberg EIT spectra until the writing of this paper confirm that the
cells have maintained original background pressure and filling levels
for almost 2 years. Figure~\ref{fig:timeline}(a) shows a
timeline of Rydberg EIT measurements collected with a single vapor cell
from this first batch (a supported version) up to the time of writing
this paper. The Rydberg state of each measurement is denoted on the
vertical axis. Each point represents a day when Rydberg EIT signals were
observed, each under different experimental conditions that prohibited a
quantitative comparison between these datasets; however, the strong EIT
signals portrayed in Figure~\ref{fig:timeline}(b-e) support the
conclusion that this cell is well sealed. The lifetime of this vapor
cell is 707 days and counting.

\subsection{EIT spectra and RF
electrometry}\label{sec:eit-spectra-and-rf-electrometry}

\begin{figure}[t]
  \centering
  \includegraphics[width=0.46\textwidth]{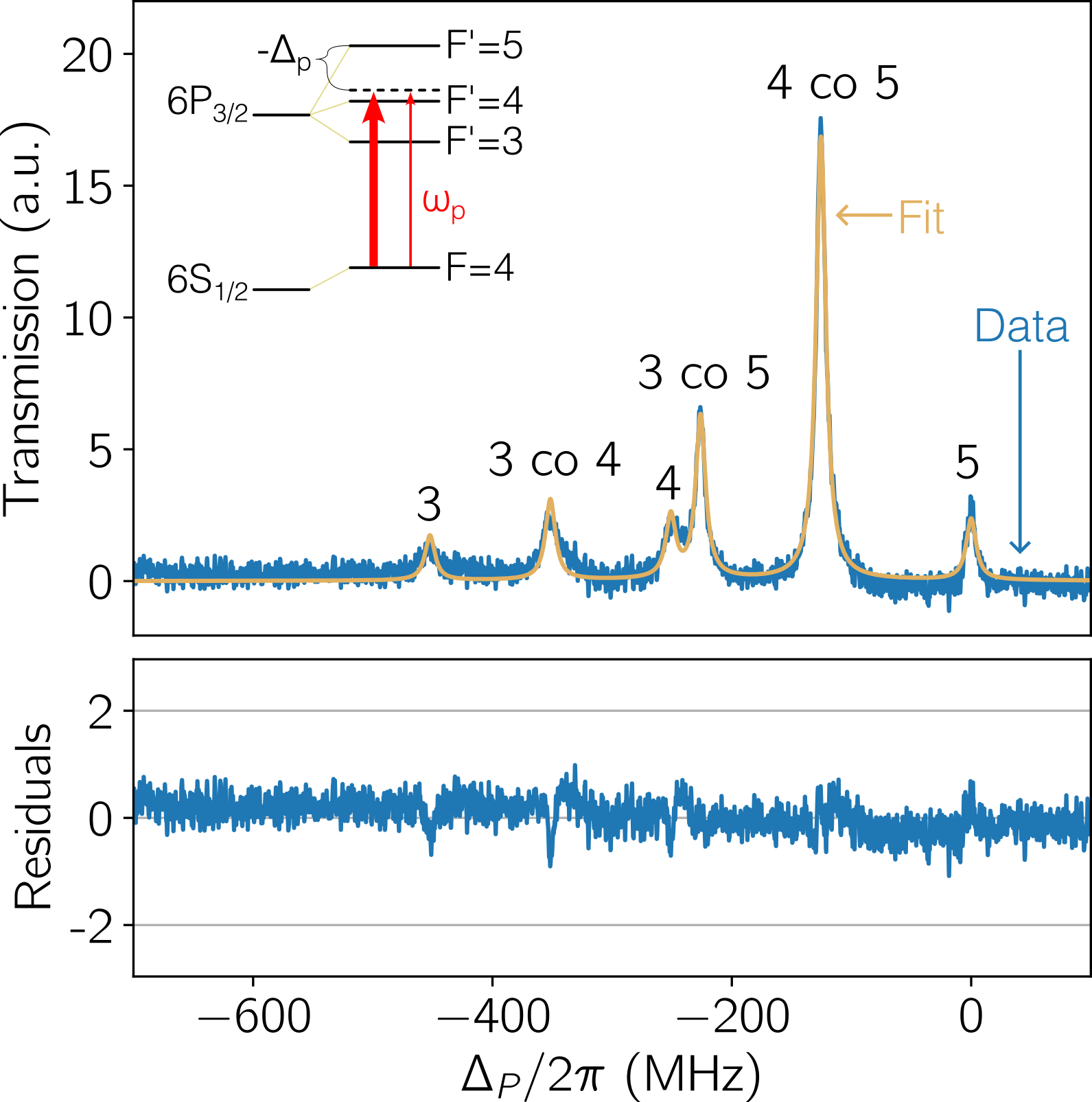}
  \caption{Saturated absorption spectrum of the Cs D2 line with the
Doppler background subtracted. The fit (Eq.~1 in
Ref.~\cite{Lei2024}) predicts the product
\(\Gamma_{vcc}\tau_{R} = 0.00(2)\), consistent with zero, confirming
adequate evacuation of the vapor cell prior to filling. Reported
uncertainty is 95\% confidence in the fit coefficient.}
  \label{fig:SAS}
\end{figure}

Before examining the EIT signals, we first characterize the atomic vapor
cell using saturated absorption spectroscopy (SAS), as shown in
Figure~\ref{fig:SAS}(a). This technique allows us to resolve the
hyperfine structure of the D2 ground-state transition. The well-defined
peaks corresponding to different hyperfine transitions validate the
cell's functionality and ensure the vapor cell was sufficiently pumped
down to minimize background pressure prior to
filling~\cite{Lei2024}. In this measurement, the probe beam power
was \(10\ \mu W\) with a beam width of \(867\ \mu m\) and the pump and
probe beams were aligned counter-propagating. A fit is applied to this
spectrum (Eq.~1 in Ref.~\cite{Lei2024}), which involves 12 free
fit parameters: six amplitudes (one for each Lamb dip peak), the
homogeneous linewidth (common to all peaks), temperature, the unitless
produce \(\Gamma_{vcc}\tau_{R}\) defined by the rate of velocity
changing collisions and the relaxation rate of the ground state, and
three parameters used to account for any nonlinearity or frequency
shifts in the frequency axis. The product
\(\Gamma_{vcc}\tau_{R} \propto p\), where \(p\) is the pressure of some
deleterious gas, is extracted from a fit to all six Lamb dips. We find
from this fit that \(\Gamma_{vcc}\tau_{R} = 0.00(2)\), consistent with
zero, which confirms adequate evacuation of the vapor cell prior to
filling. The line width of \(\Gamma = 10.03(1)\ MHz\) includes
contributions from the natural lifetime of the P state, power
broadening, and a slight broadening effect caused by a known magnetic
field \(< 0.1\ mT\) aligned with the polarization axis of the lasers.

\begin{figure}[t]
  \centering
  \includegraphics[width=\textwidth]{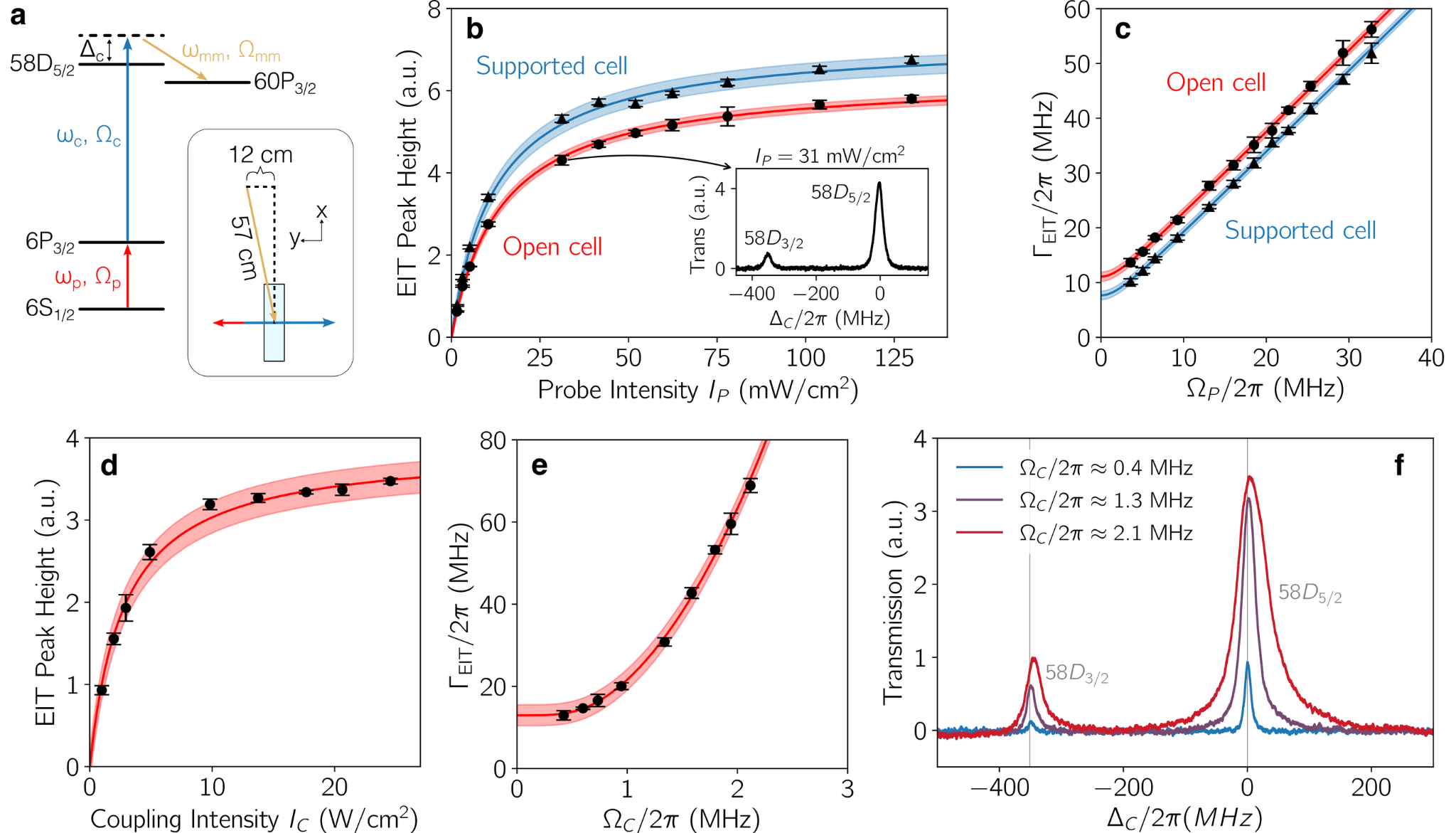}
  \caption{(a) Energy diagram corresponding with
Figures~\ref{fig:EIT} and \ref{fig:AT}. Each field
frequency (\(\omega_{i}\)), Rabi frequency (\(\Omega_{i}\)), and
detuning (\(\Delta_{i}\)) are reported in the text, where
\(i = P,\ C\ ,mm\) refer to probe, coupling, and millimeter wave fields,
respectively, and \(\Delta_{P}\)=\(\Delta_{mm} = 0\ MHz\). Inset: horn
antenna distance and angle with respect to vapor cell; all fields
linearly polarized along the z-axis. EIT peak height (b,d) and FWHM
(\(\Gamma_{EIT}\)) (c,e) as a function of probe (b,c) or coupling (d,e)
laser intensity/Rabi frequency, fits detailed in the text. (f) Measured
EIT signals at three coupling laser Rabi frequencies showing the
asymmetric broadening at high \(\Omega_{C}\). Error bars given by
\(k\sigma\), where \(k = 2.78\) (95\% confidence of 5 measurement
repeats) and \(\sigma\) is measurement standard deviation. Shaded areas
of fits denote 95\% prediction bounds.}
  \label{fig:EIT}
\end{figure}

A typical system for the absolute measurement of an RF electric field
amplitude using Rydberg Cs atoms is depicted in
Figure~\ref{fig:EIT}(a). We used this system to study the
fabricated glass vapor cells and characterize their performance, finding
that these wafer-processed cells indeed support the measurement of
millimeter wave fields and show minimal deleterious effects from
charging or stray fields within the cells.

We excite the atoms to the 58D\(_{5/2}\) Rydberg state and step either
the probe or the coupling laser power while monitoring the EIT spectrum.
In Figure~\ref{fig:EIT}, we plot the resulting EIT peak height
(b,d) and full width at half-maximum (FWHM) (c,e) as the probe (b,c) or
the coupling (d,e) laser power (reported as intensity or Rabi frequency)
increases. Given the Gaussian intensity profile of the probe laser, the
peak intensity (\(I_{P}\) or \(I_{C}\)) is given by

\begin{equation}
I_{P,C} = \frac{2P_{P,C}}{\pi w_{P,C}^{2}}\ ,
\label{eq:1}
\end{equation}

where \(P_{P,C}\) is the probe (\(P\)) or the coupling (\(C\)) power
prior to entering the vapor cell and \(w_{P} = 350\ \mu m\),
\(w_{C} = 360\ \mu m\) is the width of each Gaussian beam as it passes
through the cell. As seen in Figure~\ref{fig:EIT}(b), the EIT
peak height increases with the probe intensity and saturates. This is
determined by fitting the expression for a resonant two-state system:

\begin{equation}
h_{EIT}\left( I_{P,C} \right) = \frac{h_{\max}I_{P,C}}{\left( I_{sat} + I_{P,C} \right)}\ ,
\label{eq:2}
\end{equation}

where \(h_{\max}\) corresponds to the maximum height of the EIT peak in
arbitrary units, and \(I_{sat}\) is the saturation intensity. The fitted
values are \(h_{\max} = 7.23(4)\)~a.u., \(I_{sat} = 12.3(2)\ mW/cm^{2}\)
for the supported cell and \(h_{\max} = 6.34(5)\)~a.u. and
\(I_{sat} = 13.9(2)\ mW/cm^{2}\) for the open cell when the coupling
laser power is held fixed at \(P_{C} \approx 6\ mW\)
(\(\Omega_{C}/2\pi \approx 0.7\ MHz\)). Throughout this paper, we report
expanded uncertainties with 95\% confidence intervals. The inset within
Figure~\ref{fig:EIT}(b) displays a typical EIT signal at one
probe intensity, illustrating the characteristic transmission peak that
defines EIT. Similarly, the EIT peak height increases with the coupling
laser and saturates at an intensity of \(2.8(1)\ W/cm^{2}\) with
\(h_{\max} = 3.89(3)\)~a.u. when the probe laser power is held fixed at
\(P_{P} \approx 20\ \mu W\) (\(\Omega_{P}/2\pi \approx 9.3\ MHz\)) as
shown in Figure~\ref{fig:EIT}(d).

In Figure~\ref{fig:EIT}(c,e), the input power of the probe or
coupling laser is reported as the Rabi frequency, where

\begin{equation}
\Omega_{P,C} = \frac{\mu_{12,23}}{\hslash}\left| E_{P,C} \right|\ ,
\label{eq:3}
\end{equation}

\(\mu_{12}\) is the dipole moment of the \(|6S_{1/2}\rangle(F = 4)\)
\(\rightarrow |6P_{3/2}\rangle(F = 5)\),\cite{Steck},
\(\mu_{23}\) is the dipole moment of the
\(|6P_{3/2}\rangle(F = 5) \rightarrow |58D_{5/2}\rangle\),
\(E_{P,C} = \sqrt{2I_{P,C}/c\varepsilon_{0}}\), \(\hslash\) is the
reduced Planck constant, \(c\) is the speed of light in
vacuum~\cite{codata}, and \(\varepsilon_{0}\) is the vacuum
permittivity. The EIT linewidth (\(\Gamma_{EIT}\)), defined as the FWHM,
increases with probe Rabi frequency, showing the expected power
broadening trend which can be modeled by

\begin{equation}
\Gamma_{EIT}\left( \Omega_{P} \right) = \sqrt{A\left( \Omega_{P}/2\pi \right)^{2} + \Gamma_{0}^{2}} + \Gamma_{P}\ ,
\label{eq:4}
\end{equation}

where \(\Gamma_{0}/2\pi = 3.72\ MHz\) accounts for the lifetime of the
Rydberg state, collisions, transit time broadening, and residual Doppler
broadening (dominant). The extracted zero-power linewidth is
\(\Gamma_{P}/2\pi = 3.9(2)\ MHz\) for the supported cell and
\(\Gamma_{P}/2\pi = 7.4(2)\ MHz\) for the open cell. Similar values of
the unitless linear fit coefficient \(A\) are measured for the two cells
and agree with the expected value of 2~\cite{Citron1977}: \(2.21(5)\)
(supported) and \(2.24(6)\) (open). As the coupling laser power is
increased, the EIT linewidth again increases, but the cause in this case
is not exclusively power broadening.\cite{Lei2024}

In Figure~\ref{fig:EIT}(f), three spectra are given showing the
EIT signal at three coupling laser Rabi frequencies with the location of
the 58D\(_{3/2}\) and 58D\(_{5/2}\) Rydberg states denoted by the
vertical lines. At highest coupling Rabi frequency, the EIT signals not
only broaden but also shift slightly to the right. This is due to Stark
shifting caused by a DC electric field induced by the strong coupling
laser as it interacts with adsorbed Cs on the surface of the windows.
Each momentum substate \(m_{J}\) undergoes different amounts of shift
due to differences in polarizabilities. Furthermore, the electric field
across the region of Rydberg atoms is not likely homogeneous. The
integrated effect is broadening of the EIT signal. Given enough coupling
laser power, we expect to be able to resolve the \(m_{J}\) substates as
in~\cite{Patrick2025} even with the expected inhomogeneous field that
we integrate over the 2~mm length of the cell. The exact dependence of
this induced electric field on coupling laser power density along with
accounting for the field inhomogeneity is difficult to know. We,
therefore, choose a simple model that accounts for saturation of the
induced field (see
Section~\ref{sec:coupling-laser-dependent-broadening}) for the EIT
linewidth dependence on coupling laser Rabi frequency in
Figure~\ref{fig:EIT}(e):

\begin{equation}
\Gamma_{EIT}\left( \Omega_{C} \right) = \sqrt{B\Omega_{C}^{4} + \Gamma_{0}^{2}} + \Gamma_{C}\ ,
\label{eq:5}
\end{equation}

see Section~\ref{sec:coupling-laser-dependent-broadening} for a
more detailed justification for this model. Using this model, the
zero-power linewidth is estimated as \(\Gamma_{C} = 8.4(2)\ MHz\), which
is in part caused by power broadening from the probe laser
(\(\Omega_{P}/2\pi \approx 9.3\ MHz\)), and the power-scaling fit
coefficient is found to be \(B = 4.6(1)\) (notably larger than
\(A \approx 2\) in Eq. 4).

\begin{figure}[t]
  \centering
  \includegraphics[width=\textwidth]{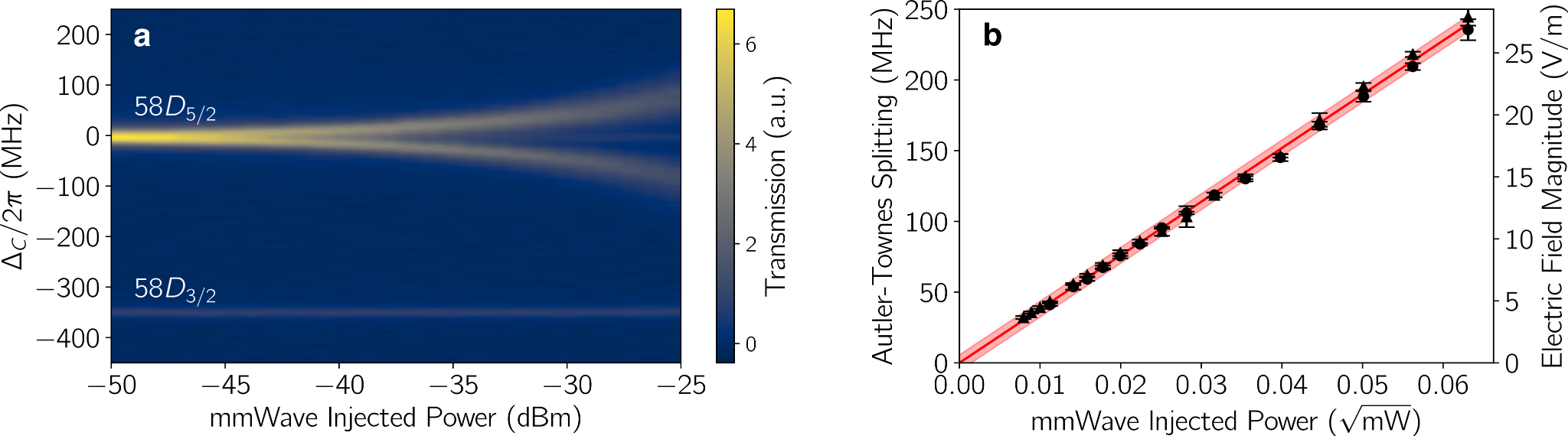}
  \caption{(a) Autler-Townes splitting of the 58D\(_{5/2}\) Rydberg state
is apparent when a 34~GHz electric field, resonant with 60P\(_{3/2}\),
is applied with increasing power fed to the horn antenna. (b) Frequency
separation between the split states (\(\Delta_{AT}\)) increases linearly
with millimeter wave field strength (\(\propto \sqrt{Power}\)). These
measurements were taken with an open cell.}
  \label{fig:AT}
\end{figure}

Next, we introduced a resonant millimeter wave field to drive
transitions between the 58D\(_{5/2}\) and 60P\(_{3/2}\) Rydberg states
of the Cs atoms to observe Autler-Townes splitting. To do this, a 34~GHz
millimeter wave signal is generated and amplified then radiated on to
the Cs vapor cell using a horn antenna positioned \(\approx 57\)~cm from
the cell at an angle of \(\approx 12^{\circ}\) from the plane of the
thin vapor cell (i.e., \(\approx 78^{\circ}\) from the axis of the
lasers, which are approximately normal to the window surfaces of the
vapor cell), see Figure~\ref{fig:EIT}(a) for a diagram of the
setup. This geometry was simply practical for our optical table setup
and not suggestive of a unique behavior of these cells. The millimeter
wave field strength is ramped by adjusting the input power defined at
the generator, allowing us to study the effect of varying millimeter
wave field amplitudes on the atomic transitions. The dielectric
structure of the cells, either open or supported, modifies the RF fields
inside the atomic volumes, which must be modeled and taken into account
in RF metrology applications. For frequencies used in this manuscript,
the cell structures are much smaller than the RF wavelength and no
resonant RF field enhancement is expected, while modest and broadband
field modulation due to cell reflection and near-field
electrostatic-limit enhancement may be present, dependent on
polarization.

The experimental results illustrating Autler-Townes (AT) splitting are
presented in Figure~\ref{fig:AT}. Figure~\ref{fig:AT}(a)
shows the transmission spectrum as a function of millimeter wave power
defined at the generator and the detuning of the coupling laser
(\(\Delta_{C}\)). As the millimeter wave power increases, a clear
splitting of the EIT resonance becomes evident. This splitting
corresponds to the Autler-Townes effect between the 58D\(_{5/2}\) and
60P\(_{3/2}\) states, confirming the strong coupling induced by the
millimeter wave field. The color scale represents the transmission
intensity in arbitrary units (a.u.), and both the splitting and the
broadening of the spectral lines become more pronounced with higher
millimeter wave power. Broadening on the AT lines at high power is
caused by inhomogeneities in the millimeter wave field within the
cell~\cite{Rotunno2023} and m\(_{J}\) sublevel splitting of the
Rydberg state. These sublevels are degenerate but become visible under a
strong applied electric field due to differences in the polarizabilities
of each state~\cite{Schlossberger2024_CPEM}.

To quantitatively analyze Autler-Townes splitting, we measure splitting
\(\Delta_{\text{AT}}\) as a function of the input millimeter wave power
to an amplifier and horn antenna, as shown in Figure
\ref{fig:AT}(b). The data exhibit a linear relationship between
\(\Delta_{\text{AT}}\) and the square root of the millimeter wave power,
which aligns with the theoretical expectation that this splitting is
equal to the Rabi frequency of the applied millimeter wave field
(\(\Omega_{mm}\)) and is proportional to the electric field amplitude
(\(E_{mm}\)) of the millimeter wave
(\(\Omega_{mm} \propto E_{mm} \propto \sqrt{P_{mm}}\), where \(P_{mm}\)
is the millimeter wave power). A linear fit to the data yields a
calibration factor of \(3797(6)\ MHz/\sqrt{mW}\). The right axis of
Figure~\ref{fig:AT}(b) gives the corresponding electric field
amplitude in volts per meter calculated using the relation

\begin{equation}
E_{mm} = \frac{\hslash}{\mu_{34}}\Delta_{\text{AT}}\ ,
\label{eq:6}
\end{equation}

where \(\Delta_{\text{AT}}\) is the measured Autler-Townes splitting,
and \(\mu_{34}\) is the transition dipole moment between the Rydberg
states. For the \(|58D_{5/2}\rangle \rightarrow |60P_{3/2}\rangle\)
transition, the dipole moment is calculated to be
\(\mu_{34} = 685 \cdot e \cdot a_{0}\)~\cite{arc}, where \(e\)
is the charge of an electron and \(a_{0}\) is the Bohr radius.

\section{Discussion}\label{discussion}

The wafer-level fabrication process presented here demonstrates an
all-glass approach to creating millimeter- to centimeter-scale vapor
cells suitable for room temperature atomic spectroscopy measurements
including Rydberg atom electrometry. The use of Borofloat
33~\cite{SchottBorofloat33}, which has a significantly lower dielectric
constant and loss tangent compared to silicon, exemplifies the potential
that this direct bonding fabrication approach offers, which is an
extending of vapor cell material options for specific application needs.

Our results confirm that these wafer-processed glass vapor cells
maintain vacuum integrity, support reproducible alkali loading, and
provide reliable Rydberg electrometry at 34~GHz. Specifically, the
observed EIT spectra are robust over several months, which confirms that
the Cs activation process adequately fills the cells and the hermetic
sealing has no apparent leak. The measured SAS and EIT linewidths are on
par with those of other microfabricated vapor
cells~\cite{Noaman2023,Zhao2023,Li2024,Giat2025_arxiv}. From the dependencies in
Figure~\ref{fig:EIT}(c) and (e) and recognizing the known Zeeman
splitting~\cite{Schlossberger2024} and photo-induced DC Stark
shifting~\cite{Patrick2025}, our analysis goes a step further than
other works by attempting to identify all sources of line broadening.
The Autler--Townes splitting measurements show a clean linear dependence
of the splitting on the square root of the incident millimeter wave
power, consistent with theoretical expectations. This linear
relationship offers a direct calibration of the electric field amplitude
at the atoms and emphasizes the suitability of these fabricated cells
for quantifying millimeter wave fields at frequencies relevant to
next-generation communication and radar systems.

There are several ways to improve and extend this platform. Although we
rely on borosilicate glass in the present work, the same direct bonding
approach can be applied to other low-loss dielectrics (assuming
ultraclean surface preparation and flatness), offering greater design
flexibility for advanced electromagnetic and optical integration. For
example, materials with even lower dielectric constants or lower RF
losses~\cite{Cano2024} could further increase measurement fidelity
in the millimeter wave bands. The ability to make structured cells
entirely out of low-loss dielectric materials opens new and exciting
possibilities for custom engineered RF-structure interactions, such as
sharp and frequency-, angle-, and polarization-specific field
enhancement inside the cell, as well as customized far-field scattering
modification. Such dielectric cells may also be better suited for high
RF power density applications. Additionally, refining the internal cell
geometry may allow for tailored mode confinement or specialized optical
paths crucial for multiaxis or high-spatial-resolution sensing.

A reaction reservoir, as is commonly found in wafer-processed vapor
cells, would also minimize the collection of alkali metal on the windows
of the cell. Given that this thin metal film is known to interact with
the coupling laser used in two-photon Rydberg EIT systems causing stray
electric fields that may broaden and shift the EIT
signal~\cite{Patrick2025}, it is beneficial to minimize the adsorption
of Cs on the surfaces of the cell where the lasers propagate.
Furthermore, an alkali getter has been shown to cause perturbations to
an incident RF field~\cite{Noaman2023}. A make-seal technique like
that presented by Maurice et al.~may even allow complete removal of the
reaction chamber~\cite{Maurice2022}. Studies of the spatial variation
in both the EIT signal of a set of Rydberg states and the AT splitting
to inform millimeter wave field uniformity across the vapor cells
discussed here are under way and will be the subject of a future
publication.

Although we demonstrate a robust batch-fabrication technique on wafer,
future research could focus on the integration of photonic components
such as waveguides, gratings, on-chip heaters, or even on-chip lasers.
Such platforms promise to reduce device footprints while enabling
wide-scale deployment in industrial, defense, and medical applications
where high-accuracy field sensing or atomic spectroscopic tools are
desired. Overall, our all-glass direct bonding process lays the
groundwork for a new generation of millimeter-scale atomic vapor cells,
providing precise Rydberg electrometry without the material constraints
posed by silicon.

We demonstrate an all-glass wafer-level fabrication process for
millimeter-scale vapor cells that enables reliable Rydberg atom
electrometry. Our results demonstrate stable vacuum-sealing and
alkali-loading processes, validated by the robust and repeatable Rydberg
EIT signals measured over many months. We characterize the EIT signal at
58D\(_{5/2}\) showing 12.0(2)~MHz minimum \emph{measured} linewidth and
develop a case for the dominant contributing factors causing this
linewidth being residual Doppler broadening and laser-induced
broadening. The dependence of the linewidth with coupling laser power
suggests that this effect may be largely due to ionization of adsorbed
Cs on the windows. Autler--Townes splittings at 34~GHz exhibits the
expected linear dependence on the square root of the input microwave
power, underscoring the suitability of these cells for precise electric
field calibration.

Looking forward, this fabrication approach paves the way for
miniaturized, low-loss vapor cells suitable for high-frequency field
sensing and offers flexibility in material choice for future systems. We
plan to explore alternative cell geometries, integrated photonic
components, internal coatings, and activation reservoirs to further
enhance long-term alkali control and minimize stray fields. Ultimately,
these improvements will broaden the applicability of Rydberg-based
sensors for radar, communication, and metrology, paving the way toward
fully integrated, mass-manufactured, miniaturized quantum devices that
operate at higher frequencies and with higher spatial resolution.

\section{Materials and Methods}\label{materials-and-methods}

\subsection{Vapor cell geometry
details}\label{sec:vapor-cell-geometry-details}

The internal volume of each vapor cell is machined using femtosecond
laser pulses and KOH etching to create either one 19~mm by 13~mm open
area reservoir or seven 19~mm long by 1.3~mm tall horizontal trenches
that are connected by a single 1~mm wide vertical trench. The total
evacuated volume of each open cell is \(494\ mm^{3}\) and the total
evacuated volume of each supported cell with trenches is
\(354.2\ mm^{3}\). This trench geometry minimizes bowing of the vapor
cell windows to facilitate multiple reflections of the lasers within the
cell over long path lengths without alignment errors (see
Section~\ref{sec:cell-height-profile} for detailed discussion of
this vapor cell geometry).

\subsection{Atomic spectroscopy
measurements}\label{sec:atomic-spectroscopy-measurements}

In this system, the ground state, \(|6S_{1/2}\rangle(F = 4)\), is
coupled to the intermediate state, \(|6P_{3/2}\rangle(F' = 5)\), via a
weak probe laser at 852.3~nm. Simultaneously, the transition between the
intermediate state and the Rydberg state, \(|58D_{5/2}\rangle\), is
driven by a strong coupling laser field at 509~nm. The probe laser is
generated using a grating-stabilized tunable single-mode laser, while
the coupling laser is produced by a frequency-doubled tunable diode
laser system capable of delivering up to 300~mW of power. The probe
laser is power stabilized with an electronically controlled variable
optical attenuator (EVOA), and a pair of acousto-optic modulators (AOMs)
are used to power stabilize and modulate the coupling laser beam at a
frequency of 37.2~kHz. The lasers are aligned to counterpropagate
through the Cs vapor cell at room temperature with the lasers
approximately normally incident on the cell windows. The probe laser
beam width is \(350\ \mu m\) and the coupling laser beam width is
\(360\ \mu m\) through the 2~mm vapor cell length. The EIT signal is
recorded through differential detection, incorporating a balanced
photodetector to minimize common noise in the probe laser with a
reference beam that also passes through the vapor cell. The detected
signal is then fed into a lock-in amplifier, synchronized with the AOM
modulation, to extract the EIT signals with improved signal-to-noise
ratio. The lock-in amplifier is configured with a time constant of
30~\(\mu\)s, a sensitivity of 200~mV (typical), and 24~dB setting for
the low pass filter roll-off. Additionally, the frequency of the probe
laser is stabilized using a reference ultra-low expansion Fabry-Perot
cavity.

To capture the EIT spectra reported in Figures~\ref{fig:EIT} and
\ref{fig:AT}, we scan the frequency of the coupling laser over
the 58D\(_{3/2}\) and 58D\(_{5/2}\) Rydberg states, observing increased
probe laser transmission when the coupling laser is near resonance with
each state. The calculated frequency difference between the
58D\(_{3/2}\) and 58D\(_{5/2}\) states is
351.21~MHz~\cite{arc} and sets the frequency axis of the laser
scan. In this experiment, the RF signal is generated by a signal
generator (up to 40~GHz), which provides a stable continuous wave
microwave output at 34.009~GHz. The output of the signal generator is
routed into a traveling wave tube amplifier (TWT), which delivers the
high gain and power necessary to drive AT splitting of the Rydberg
state. A WR-28 rectangular waveguide gain horn efficiently couples the
amplified microwave signal into free space to illuminate the vapor cell.
The combined TWT gain and coax cable and connector loss between the
signal generator output connector and the coax connector at the input of
the horn antenna is \(\approx\)39.5~dB and the specified antenna gain is
23~dBi.

\subsection{Coupling laser dependent
broadening}\label{sec:coupling-laser-dependent-broadening}

The dependence of the EIT linewidth on the coupling laser power is
challenging to model analytically given an inhomogeneous DC electric
field within the cell the scales with the coupling laser power density
on the cell window. As demonstrated by Patrick, et
al.~\cite{Patrick2025}, the field within the vapor cell exhibits a
saturating trend with increasing coupling laser power on the cell
surface. We attribute this behavior to the photoelectric effect, where
the incident laser generates a charge density on the cell windows,
resulting in a non-uniform field across the optical length of the vapor
cell. The number of photons in the laser is \(n = P/(hf)\), where \(P\)
is the laser power, \(h\) is the Planck constant, and \(f\) is the laser
frequency. At low photon numbers, the number of charges generated is
proportional to the number of photons hitting the glass per unit time,
where we introduce an efficiency and recombination rate scaling factor
(\(\eta\)). Given the observed saturation of the field (i.e., the
generated charges), the total charge generated is then given by
\(n_{q} = \eta n/\left( 1 + n/n_{sat} \right)\). The electric field
emitted by the surface charges is estimated as
\(E = k/r^{2}\int\sigma dA = n_{q}q/4\pi\varepsilon_{0}r^{2}\).
Substituting the expressions for \(n_{q}\) and \(n\), we obtain

\begin{equation}
E = \frac{qP\eta}{4\pi\varepsilon_{0}r^{2}hf\left( 1 + P/P_{sat} \right)}\ .
\label{eq:7}
\end{equation}

The Stark effect induces a shift, \(\Delta_{s} = \alpha|E|^{2}/4\),
which leads to broadening due to the variation of the displacement
vector magnitude (\(r\)) within the vapor cell for the set of Rydberg
atoms probed. To first order, we assume symmetry between the entrance
and exit windows. By substituting the Rabi rate for the optical field,
we derive the average shift across the vapor cell as

\begin{equation}
\Delta_{avg} = - \frac{\alpha\eta^{2}q^{2}c^{2}w_{C}^{4}h^{2}}{3072f^{2}\mu^{4}r^{3}}|_{r_{0}}^{r'} \cdot \left( \frac{\Omega_{C}^{2}}{1 + \Omega_{C}^{2}/\Omega_{sat}^{2}} \right)^{2} = C_{1} \cdot \left( \frac{\Omega_{C}^{2}}{1 + \Omega_{C}^{2}/\Omega_{sat}^{2}} \right)^{2}\ ,
\label{eq:8}
\end{equation}

where \(C_{1}\) encompasses the constants, quantum efficiency, and
non-uniformity. The total effect of the coupling laser on the line shape
is then described by

\begin{equation}
\Gamma_{EIT}\left( \Omega_{C} \right) = \sqrt{(C_{1}\frac{\Omega_{C}^{4}}{\left( 1 + \Omega_{C}^{2}/\Omega_{sat}^{2} \right)^{2}})^{2} + C_{2}\left( \Omega_{C}/2\pi \right)^{2} + \Gamma_{0}^{2}}\ ,
\label{eq:9}
\end{equation}

where \(C_{2} \ll C_{1}^{2}\) is a fit coefficient associated with the
coupling laser power broadening term that can be ignored for
\(\Omega_{C}/2\pi < 15\ MHz\) at the Rydberg state probed in these
measurements. This model predicts an \(\Omega_{C}^{8}\) scaling inside
the square root in the linear regime and an \(\Omega_{C}^{4}\) scaling
near saturation. The coupling laser power sweep study reported in
Figure~\ref{fig:EIT}(d-f) is near the saturation power
associated with the saturation of the photo-induced electric field such
that measured EIT linewidths fit a model that scales as
\(\Omega_{C}^{4}\) inside the square root.

\begin{figure}[t]
  \centering
  \includegraphics[width=\textwidth]{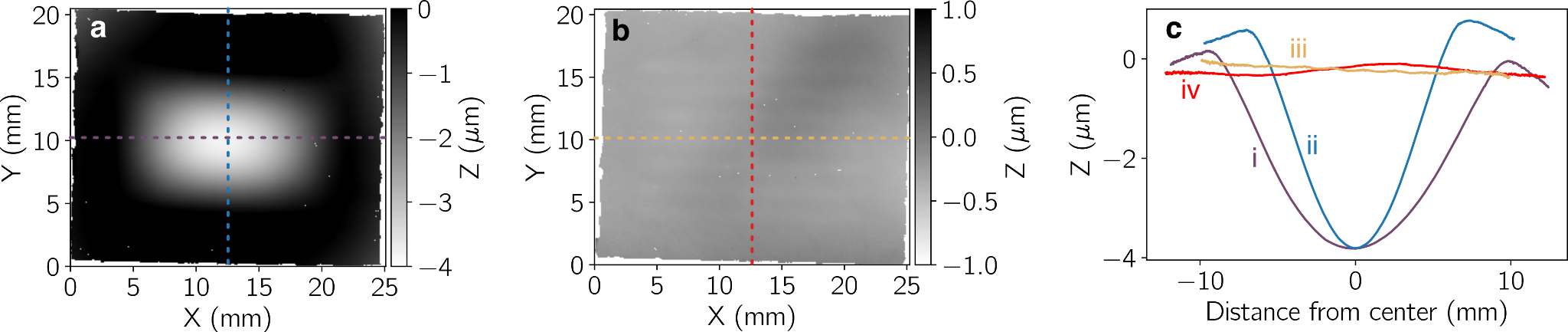}
  \caption{(a) Height profile of one vapor cell window without glass
supports (the open cell). (b) Height profile of one supported cell. (c)
Slices along the X axis (i and iii) and along the Y axis (ii and iv)
through the center of each cell - i and ii correspond to (a), the open
cell without supports, and iii and iv correspond to (b), the cell with
supports, as indicated by the colored dashed lines.}
  \label{fig:CellHeightProfile}
\end{figure}

\subsection{Explanation of Cell
Geometry}\label{sec:cell-height-profile}

The specific vapor cell geometry is chosen to enable the construction of
a sensor array. In that system, the coupling laser is internally
reflected within the vapor cell along the long (X) axis to overlap a
one-dimensional array of probe lasers. This internal reflection
technique maximizes the power efficiency of the coupling laser; however,
alignment of the two lasers is exceptionally susceptible to any
non-planarity of the vapor cell windows supporting the mirror coatings.

The pressure difference between the evacuated volume of the vapor cell
and atmosphere causes the windows of the cell to bow inwards,
Figure~\ref{fig:CellHeightProfile}(a). The dimensions of the
evacuated region in this open, non-supported cell is 13~mm tall (Y) by
19~mm wide (X) by 2~mm thick (Z). The maximum deflection of the outer
surface of one window is measured to be about \(3.8\ \mu m\). Given this
deflection, we estimate that the position error of the internally
reflected coupling beam in our desired architecture would be greater
than \(130\ \mu m\) over the 19~mm cell length given the
9.45\(^{\circ}\) designed incidence angle. This position error is nearly
a beam width meaning that array elements at the end of each row would
have weak to no signal due to severe misalignment of the two
counter-propagating lasers. On the other hand, the window deflection of
the vapor cell design with glass support structures
(Figure~\ref{fig:CellHeightProfile}(b)) is \(\ll 1\ \mu m\) such
that any alignment error caused by curvature in the windows is
negligible. The maximum deflection within an open trench within the
supported cell is at the edge of the tool resolution and is estimated to
be less than \(100\ nm\) from the height profile data shown in
Figure~\ref{fig:CellHeightProfile}(b).

\textbf{Data availability.} All of the data presented in this paper and
used to support the conclusions of this article is published under the
identifier doi:10.18434/mds2-3762.

\textbf{Acknowledgments.} The authors thank Mike Labella and Meixu Bao
of Penn State University for their help with the fabrication of the
vapor cells. The authors additionally thank John Kitching and Gabriela
Martinez of NIST for their instructive guidance on laser-heated
activation of alkali metal dispensers and Lingyun Chai and Daniel
Hammerland for technical reviews of this manuscript. The research was
partially funded by the National Institute of Standards and Technology
Innovations in Measurement Sciences (IMS) Program and the Materials
Research Institute at Penn State University.

\textbf{Author contributions.} A.B.A., V.A., and D.L. conceived the
research and supervised all aspects of the project. All authors
participated in designing the experiment, analyzing the optical results,
and drafting the paper. H.S., G.L., M.L., C.E., and D.L. developed the
wafer-level vapor cell fabrication process and manufactured the reported
glass cavities. A.B.A. and A.M. performed theoretical analysis and
optical measurements. M.S., G.H., and C.H. participated in related
discussions.

\textbf{Conflict of Interest.} The authors declare no conflict of
interest.

\bibliography{main_bib}

\end{document}